\documentstyle[aps,epsfig,subfigure]{revtex}

\begin{document}
\title{Fluctuations of topological disclination lines in nematics: 
       renormalization of the string model} 
\author{D. Sven\v sek\footnote{Author for correspondence. 
Phone: +386 1 4766587, fax: +386 1 2517281,\\
e-mail: {\tt daniel@fiz.uni-lj.si}}
and S. \v Zumer}
\address{Department of Physics, University of Ljubljana,\\ 
         Jadranska 19, SI-1000 Ljubljana, Slovenia}
\date{\today}
\draft

\maketitle	

\begin{abstract}
The fluctuation eigenmode problem of the nematic topological disclination line 
with strength $\pm 1/2$ is solved for the complete nematic tensor order parameter.
The line tension concept of a defect line is assessed, the line tension is properly defined.
Exact relaxation rates and thermal amplitudes of the fluctuations are determined.
It is shown that within the simple string model of the defect line the amplitude of its 
thermal
fluctuations is significantly underestimated due to the neglect of higher radial modes.
The extent of universality of the results concerning other systems possessing line defects
is discussed.

\end{abstract}


The fluctuation problem of the nematic disclination line is one of the central yet 
untackled problems 
in the field of liquid crystals, despite the fact that
topological disclination lines
are the most distinct and regularly observed characteristic 
of the nematic phase 
\cite{chuang91,ishikawa,wang98,thiberge}
and are reflected even in its name.  
Lately they have been playing a role as experimental objects in studies
related to cosmology \cite{chuang,bowick,kibble}.
Controlled experiments with isolated lines or pairs of lines can be performed 
to study their motion \cite{cladis86,bogi}.
Thermal fluctuations of disclination lines can be
observed and analyzed simply by polarization microscopy \cite{mertelj1,mertelj2}.

\begin{figure}
\begin{center}
	\epsfig{file=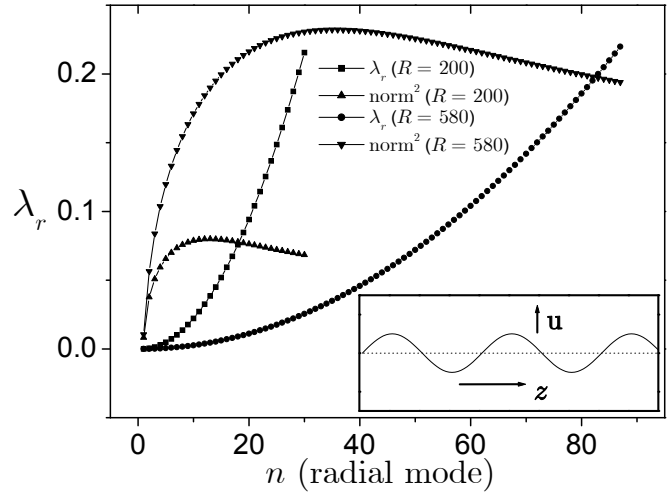,height=7cm}
\end{center}
\caption{Radial eigenvalue $\lambda_r$ (in units of $1/\tau$) 
         and the norm squared of the radial eigenfunctions (arbitrary units) for two
	 radii of confinement. 
	 For $R=200$, $\lambda_r^{(1)}=1.4\cdot 10^{-5}$, 
	              $\lambda_r^{(2)}=4.5\cdot 10^{-4}$,
		      $\lambda_r^{(3)}=1.4\cdot 10^{-3}$; 
         note that $\lambda_r^{(2)}/\lambda_r^{(1)}\approx 30$!
	 Inset: the string fluctuation of the disclination line.}
\label{lamnorsin}
\end{figure}

In this paper, we present the solution of the full tensorial fluctuation problem of a straight
wedge disclination line with strength $\pm 1/2$, in terms of finding the exact 
eigenmodes. We focus on the soft modes that correspond to string fluctuations 
(Fig.~\ref{lamnorsin}, inset)
and challenge the validity of the string model of the disclination line \cite{degennesbook}. 
The main results of our study are not limited to nematics,
as our description does not include the physical system-specific elastic anisotropy or 
hydrodynamic flow.
In particular, 
they do not depend on the winding number of the defect and can be directly applied to
disclination lines in smectics-C and vortex lines in superfluids.
The fluctuations of the strength $1$ nematic disclination line were studied 
in the director description
\cite{ziherlw,ziherlwo}, precluding the string fluctuations.
Moreover, the topological $\pm 1/2$ disclinations  
cannot be described by the Frank director elasticity, whereas the integer disclinations 
are known to be unstable.

The director free energy per unit length --- line tension --- of a 
straight disclination line with strength $s$ is
\begin{equation}
	{\cal F}_0 = \pi s^2 K \ln{R\over r_0},\label{F0}
\end{equation}
where $K$ is the Frank elastic constant,
$R$ the system size and $r_0$ a microscopic cutoff. Hence, in a model the 
disclination line can be considered as a simple string under tension \cite[p. 178]{degennesbook}. 
The energy cost of its (overdamped) fluctuation modes (Fig.~\ref{lamnorsin}, inset), 
which can be attributed 
to the increase in length,
$\Delta l = \int\!dz (\partial u/\partial z)^2/2$,
is thus
\begin{equation}
	\Delta F={1\over 2}k^2 u^2 {\cal F}_0\int\!\!dz \cos^2 kz,\label{DF}
\end{equation}
where $u$ is the amplitude and $k$ the wave vector of the mode.
The string model of a
line defect is general \cite{sonin,geminard,colson} 
and not restricted to nematic disclination lines.

Let us set the scene for the complete description of the disclination line in terms of the 
nematic tensor order parameter $\sf Q$.
Cylindrical coordinates $(r,\phi,z)$ with corresponding $\bf r$-dependent 
orthonormal base vectors
$\{{\bf\hat{e}}_r,{\bf\hat{e}}_\phi,{\bf\hat{e}}_z\}$
will be used. The unperturbed disclination line (the ground state, as referred to from now on) 
coincides with the $z$ axis. 
The ground state is $z$-independent, hence, the $z$-dependence of the eigenmodes  
is sinusoidal.
In one (elastic) constant approximation, the free energy density $f({\sf Q},\nabla{\sf Q})$
reads
\begin{equation}
	f = \textstyle{1\over 2}A\, {\rm Tr}{\sf Q}^2+
	    \textstyle{1\over 3}B\, {\rm Tr}{\sf Q}^3+
	    \textstyle{1\over 4}C\, ({\rm Tr}{\sf Q}^2)^2+
	    \textstyle{1\over 2}L\, {\rm Tr}(\nabla{\sf Q}\cdot\nabla{\sf Q})\label{f}
\end{equation}
and is invariant upon
a homogeneous rotation of the $\sf Q$-tensor.
Hence, the $\sf Q$-eigensystem rotates as
$\psi = \psi_0+(s-1)\phi$ with respect to the above base vectors 
when we encircle a defect of strength $s$ located at the origin;
$\psi$ is the angle between ${\bf\hat{e}}_r$ and the tensor axis corresponding to 
the director in the far-field, while
$\psi_0$ is the free parameter of the defect configuration,
corresponding to the angle between the director axis at $\phi=0$ and the $x$ axis.
There is no dependence on $\phi$ other than this rotation, i.e., the scalar invariants
of $\sf Q$ (the degree of order and biaxiality) are $\phi$-independent ---  a
``generalized'' cylindrical symmetry.

Defining another $\bf r$-dependent orthonormal triad 
$\{{\bf\hat{e}}_1,{\bf\hat{e}}_2,{\bf\hat{e}}_z\}$, where 
${\bf\hat{e}}_1 = {\bf\hat{e}}_r\cos\psi+{\bf\hat{e}}_\phi\sin\psi$ and
${\bf\hat{e}}_2 = -{\bf\hat{e}}_r\sin\psi+{\bf\hat{e}}_\phi\cos\psi$,
in the unperturbed configuration 
the $\sf Q$-tensor eigensystem coincides with this triad everywhere.
Defining further the orthonormal 
symmetric traceless base tensors \cite{hess75}: 
${\sf T}_0 = (3{\bf\hat{e}}_z\otimes{\bf\hat{e}}_z - {\sf I})/\sqrt{6}$,
${\sf T}_1 = ({\bf\hat{e}}_1\otimes{\bf\hat{e}}_1 -
                        {\bf\hat{e}}_2\otimes{\bf\hat{e}}_2)/\sqrt{2}$,
${\sf T}_{-1} = ({\bf\hat{e}}_1\otimes{\bf\hat{e}}_2 +
                        {\bf\hat{e}}_2\otimes{\bf\hat{e}}_1)/\sqrt{2}$,
${\sf T}_{2} = ({\bf\hat{e}}_z\otimes{\bf\hat{e}}_1 +
                        {\bf\hat{e}}_1\otimes{\bf\hat{e}}_z)/\sqrt{2}$,
${\sf T}_{-2} = ({\bf\hat{e}}_z\otimes{\bf\hat{e}}_2 +
                        {\bf\hat{e}}_2\otimes{\bf\hat{e}}_z)/\sqrt{2}$,
the ground state $\sf Q$-tensor 
components are $\phi$-independent, while the components of an arbitrary perturbation 
can be translated in $\phi$ not changing the energy. 
Hence, the $\phi$-dependence of the eigenmode components is sinusoidal.

In the ocean of fluctuations one should first look for the Goldstone modes, as they yield
families of slowly relaxing and spatially extensive ``soft'' fluctuation modes 
that can be easily observed.
``Massive'' fluctuations are less interesting, as they are short-living ($\approx$100~ns)
and localized ($\approx$10~nm).
Besides the homogeneous rotations of the $\sf Q$-tensor, which in a deformed nematic
are soft only in the
one constant approximation, in a deformed system there exist additional nontrivial
Goldstone modes,
in our case the interest is in those corresponding to the displacement 
of the disclination line. 
Since translational degrees of
freedom are absent in the analysis, the ``displacement'' is achieved by modifying 
(perturbing) the order parameter field. Modulating the displacement modes sinusoidally
along $z$ results in the string fluctuations of the disclination line
(Fig.~\ref{lamnorsin}, inset).

Expressing the $\sf Q$-tensor as
${\sf Q}({\bf r},t) = q_i({\bf r},t) {\sf T}_i({\bf r})$,
the elastic (gradient) part of (\ref{f}) is
\begin{eqnarray}
	f^e &=&
	    \textstyle{L\over 2} \left\{
	    \left({\partial q_i\over\partial r}\right)^2+
	    \left({\partial q_i\over\partial z}\right)^2+\right.\label{elast} \\
	    &+&\left.\textstyle{1\over r^2}\left[
	    \left({\partial q_0\over\partial\phi}\right)^2+
	    \left({\partial q_1\over\partial\phi}-2s\, q_{-1}\right)^2+
	    \left({\partial q_{-1}\over\partial\phi}+2s\, q_1\right)^2+
	    \left({\partial q_2\over\partial\phi}-s\, q_{-2}\right)^2+
	    \left({\partial q_{-2}\over\partial\phi}+s\, q_2\right)^2
	    \right]
	    \right\}\nonumber,
\end{eqnarray}
while the homogeneous (gradient-independent) part of (\ref{f}) is a polynomial in $q_i$.
We introduce dimensionless quantities
$r\gets r/\xi$, $t\gets t/\tau$, $(A,B,C)\gets (A,B,C)\xi^2/L$,
with the correlation length of the degree of order $\xi$, typically a few nm,
and the characteristic time $\tau=\mu_1\xi^2/L$, typically tens of ns, 
where $\mu_1$ is the bare rotational viscosity, i.e., $\gamma_1=9S^2\mu_1/2$; $\gamma_1$ is
the usual rotational viscosity and $S$ the degree of order.
Neglecting the hydrodynamic flow, the order parameter dynamics is governed 
by the time-dependent Ginzburg-Landau equation 
\begin{equation}
        \dot{q}_i = 
	\nabla\cdot{\partial f\over\partial\nabla q_i} - {\partial f\over\partial q_i},
	\label{tdgl}	
\end{equation}
where the time derivative vanishes in the ground state.

Owing to the generalized cylindrical symmetry, the ground state involves only
${\sf Q}_0({\bf r})=a_0(r){\sf T}_0+a_1(r){\sf T}_1$,
as opposed to perturbations:
$\Delta{\sf Q}({\bf r},t) = x_i({\bf r},t){\sf T}_i$.
The fluctuation eigenmodes $x_i$, satisfying $\dot{x}_i=-\lambda x_i$,
are sought by the ans\"{a}tze
\begin{eqnarray}
	x_i &=& R_{i,m}(r)\,\Phi_i(m\phi)\, \sin(kz)\, \exp(-\lambda t),\quad i = 0,\pm 1, 
	\label{ansatz1} \\
	x_i &=& R_{i,n}(r)\,\Phi_i(n\phi)\,\sin(kz)\,\exp(-\lambda t),\quad i = \pm 2, 
	\label{ansatz2}
\end{eqnarray}
where $\Phi_i=\cos$ for $i\ge 0$ and $\Phi_i=\sin$ for $i<0$
(the global angular phase and the $z$ phase are arbitrary),
$m$ is an integer, whereas $n=1/2,3/2,5/2,...$ is a half-integer due to the
continuity and differentiability requirements (spinor symmetry of the base tensors)!
In the one constant approximation, the sets of components (\ref{ansatz1})
and (\ref{ansatz2}) are
not coupled, i.e., in-plane and out-of-plane 
fluctuations are independent. Furthermore, the $z$-dependence is fully decoupled, i.e., 
the eigenfunctions $R_i(r)$ do not depend on $k$ and the cross-section structure 
of the disclination line is not affected by the $z$ modulation.

We focus on the in-plane fluctuations.
Putting the ansatz (\ref{ansatz1}) into Eq.~(\ref{tdgl}),
the eigensystem for the radial functions $R_{i,m}(r)$ remains, 
with $\lambda=\lambda_r+k^2$, where $\lambda_r$ is the radial eigenvalue:
\begin{eqnarray}
	&&\nabla^2 R_{0,m} + \left(\lambda_r-g_{0}(r)-{m^2\over r^2}\right)R_{0,m} + 
	               g_{01}(r)\,R_{1,m} = 0,\label{R0}\\
	&&\nabla^2 R_{1,m} + \left(\lambda_r-g_{1}(r)-{m^2\!\!+\!4s^2\over r^2}\right)R_{1,m} -
	{4sm\over r^2}\,R_{-1,m} + g_{01}(r)\,R_{0,m} = 0,\label{R1}\\
	&&\nabla^2 R_{-1,m} + \left(\lambda_r-g_{-1}(r)-{m^2\!\!+\!4s^2\over r^2}\right)R_{-1,m} -
	{4sm\over r^2}\,R_{1,m} = 0;\label{R-1}
\end{eqnarray}
the $g$'s are quadratic polynomials of the ground state components.
Note that the 
operator (\ref{R0})--(\ref{R-1}) is self-adjoint.
Also note that defects with strengths $s$ 
and $-s$ are formally equivalent, i.e.,
changing the sign of the defect and redefining
${\sf T}_{\{-1,-2\}}\to -{\sf T}_{\{-1,-2\}}$ conserves the sign of $s$ in
the equations.

For an infinite system, one can determine the homogeneous displacement 
mode directly by construction. 
Putting ${\sf Q}_0({\bf r}-{\bf u})={\sf Q}_0({\bf r})+\Delta{\sf Q}({\bf r})$,
the perturbation $\Delta{\sf Q}$ corresponding to the displacement $\bf u$ is
${\Delta{\sf Q}({\bf r})} = -{\bf u}\cdot{\partial{\sf Q}_0/\partial{\bf r}}$, which reads
\begin{equation}
	x_0({\bf r}){\sf T}_0 + x_1({\bf r}){\sf T}_1 + x_{-1}({\bf r}){\sf T}_{-1}=  
	  -{\bf u}\cdot\left(
	\hat{\bf e}_r{\partial a_0\over\partial r}{\sf T}_0 +
		\hat{\bf e}_r{\partial a_1\over\partial r}{\sf T}_1 +
		\hat{\bf e}_\phi\,2s{a_1\over r}{\sf T}_{-1} \right).\label{displac}
\end{equation}
Thus, $x_i$ in (\ref{displac}) are the eigenfunctions corresponding to $\lambda_r=0$ and $m=1$. 
The lowest family of string modes is generated by adding the $z$-dependence. Hence,
$\lambda=k^2$, or $\lambda=L k^2/\mu_1=K k^2/\gamma_1$ in physical units
($K=9S^2 L/2$),
which is thus the exact result for the relaxation rate of the string mode, not obtainable 
in the director picture.

The energy cost of an arbitrary fluctuation eigenmode is given by
$\Delta F({\bf x})=\lambda\int\!\!dV x_i^2/2$.
For the homogeneous displacement mode (\ref{displac}), 
integrating over $\phi$, this results in
\begin{equation}
	\Delta F={1\over 2}k^2 u^2 \underbrace{\pi\int\!\!r dr\left[
	\left({\partial a_0\over\partial r}\right)^2+
	\left({\partial a_1\over\partial r}\right)^2+4s^2\left({a_1\over r}\right)^2
	\right]}\int\!\!dz \cos^2 kz.\label{deltaF}
\end{equation}
Comparing (\ref{deltaF}) (multiplied by $L\xi$ to revert to physical units) 
with the form of the energy cost (\ref{DF}) of the simple string
fluctuation, a line tension can be defined for the disclination line, as indicated
by the underbrace. 
If the string model was accurate, the line tension defined in (\ref{deltaF}) would be the
actual free energy (\ref{f}) of the unperturbed disclination line per unit length.
One verifies that it is exactly the elastic free energy (\ref{elast}) per unit length, 
whereas the homogeneous contributions are absent. 
Only the elastic terms contribute to the line tension!
This is a general statement --- any other homogeneous contributions, should they exist, e.g., 
electric, do not enter the line tension. 
It must be stressed that within the isotropic order parameter elasticity this finding is 
universal.
Physically it means that 
the string fluctuation conserves the volume of the defect line, as opposed to 
the fluctuation of a real string.
Thus, far from the defect core, where the homogeneous free energy 
contributions vanish, the string model is exact. Moreover, using the bulk values
of $a_0$ and $a_1$, the line tension (\ref{F0}) is recovered exactly.
Due to the director
distortion, however, this regime is not approached exponentially but by a power-law.

\begin{figure}
\begin{center}
	\epsfig{file=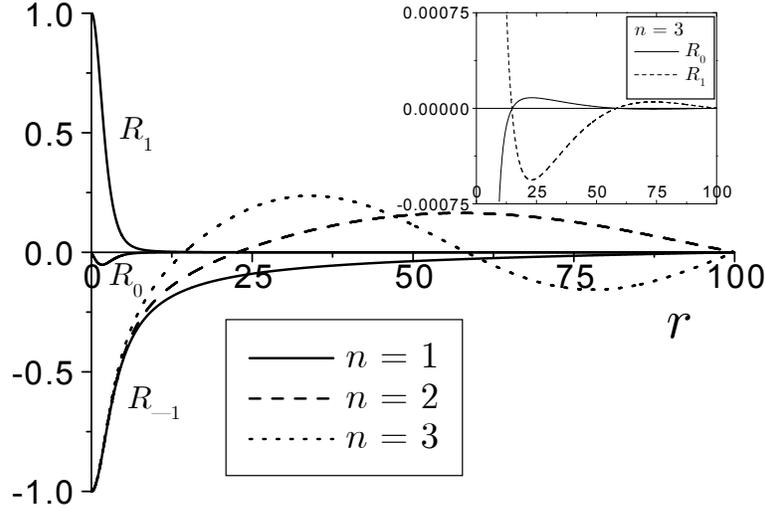,height=7cm}
\end{center}
\caption{Three lowest radial eigenfunctions (not normalized) for $m=1$. 
         Functions $R_{0,1,n}$ and $R_{1,1,n}$ 
         overlap on the scale of the big graph; the detail for $n=3$ is shown in the inset ---
         note the power-law tail and the zeros, which coincide with the zeros of $R_{-1}$.}
\label{modeki}
\end{figure}

Let us now focus on the $m=1$ fluctuations with $\lambda_r>0$. 
Unlike $R_{-1}$, the functions $R_0$ and $R_1$ are ``localized'' (Fig.~\ref{modeki}), 
since $g_0$, $g_1\to$ const.~$>0$ (of order 1), 
while $g_{-1}\to -4s^2/r^2$.
Asimptotically, $R_{-1,m}(r)$ behaves as a combination of Bessel functions
$J_m(\sqrt{\lambda_r}r)$ and $Y_m(\sqrt{\lambda_r}r)$, whereas
$R_{\{0,1\}}$ first decay as $\exp(-r)/\sqrt{r}$, 
followed by a power-law asymptotics (Fig.~\ref{modeki}, inset):
$R_{\{0,1\},m}\propto 4 m s\, R_{-1,m}/r^2$.
In solving (\ref{R0})--(\ref{R-1}), we confine the 
system at a radius $R$ with the restriction $R_{i,m,n}(R)=0$ in order to get a 
discrete spectrum $\lambda_r^{(n)}$, $n=1,2,3,...$
One might argue that this boundary condition is rather unphysical. In a real sample, however, 
one never deals with an ideally isolated disclination line --- the observed line is surrounded
by other defects and irregularities. The boundary condition should therefore be viewed as due
to an effective confinement. In the limit $R\to\infty$, physical observables should be 
only weakly dependent of $R$ anyhow, e.g., like the free energy (\ref{F0}).
Furthermore, if one uses the boundary condition $R_i'(R)=0$, the lowest eigenmode is of
the growing type, $\lambda_r<0$, where $\lambda_r\propto 1/R^2$, 
reflecting the instability of the defect towards the escape from the system.
This finite-size effect is often manifest in experimental situations.

The boundary conditions at $r=0$ are obtained by finding the analytic behavior of the
ground state and the perturbations for small $r$:
$a_0' \to 0$, 
$a_1 \to {\rm const.}\times r^{\vert 2s\vert}$,
$R_{0,m}\to r^m$, and
$R_{\pm 1,m}\to \pm a_1\, r^{\vert m-2s\vert} + a_2\, r^{\vert m+2s\vert}$.
The eigensystem (\ref{R0})--(\ref{R-1}) is discretized and efficiently solved by a 
multidimensional Newton relaxation method \cite[p. 588]{recipes}, Fig.~\ref{modeki}. 
In the relevant limit $R\gg\xi$,
the radial dependence of the lowest modes is identical for $r\ll R$ 
and given by (\ref{displac}), 
except for the normalization, i.e., each of these modes
contributes to the displacement $\bf u$ of the central part! 
Still in the limit
$R\gg\xi$, the difference $\sqrt{\lambda_r^{(n+1)}}-\sqrt{\lambda_r^{(n)}}$ 
approaches $\pi/R$ 
when going to higher modes, i.e., $\lambda_r^{(n)}\to ({\rm const.}+ n\pi/R)^2$, 
Fig.~\ref{lamnorsin}.
Similarly, for $n\gg 1$, the eigenvalue of the $n$th mode
scales as $\lambda_r^{(n)}\propto 1/R^2$ with the system size $R$.

\begin{figure}
\begin{center}
	\vbox{\subfigure[$k=\pi/100$]{\epsfig{file=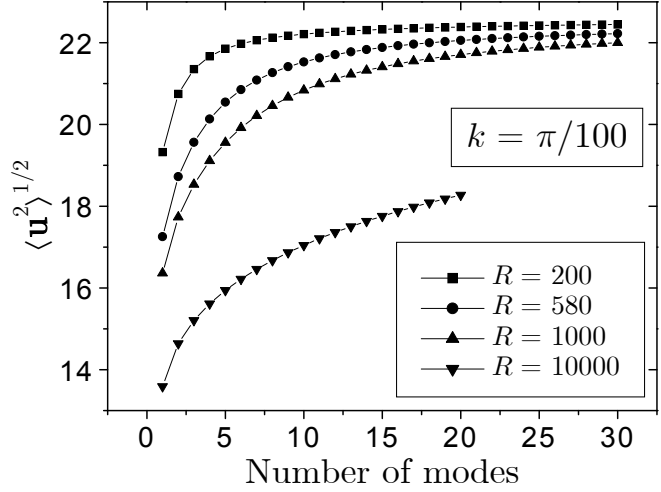,height=7cm}}
	      \subfigure[$k=\pi/1000$]{\epsfig{file=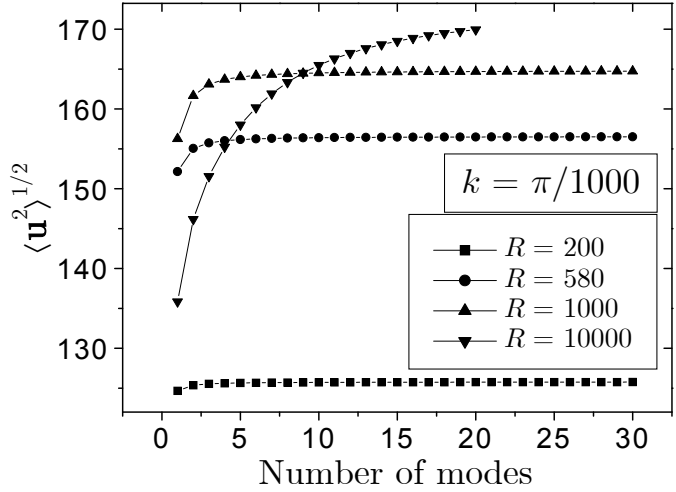,height=7cm}}}
\end{center}
\caption{rms displacement of the disclination line (the length unit is $\xi$) 
         vs. the number of 
         radial modes summed in the regimes of a) weak and b) severe confinement; 
	 $n=1$ corresponds to the non-renormalized string model.
         The maximum size of the computational domain is $R=580$, the values
         beyond this range ($R=1000$, $R=10000$) are extrapolated.}
\label{amplituda}
\end{figure}

We have shown that within the simple string model the degrees of freedom of the
disclination are cut to the lowest $m=1$ mode only. Thus, quite universally, 
the contributions of the higher
modes represent a renormalization of the model. 
Thermal fluctuation amplitude at $r=0$ of the $k$th Fourier component ($z$ direction) is
\begin{equation}
	\langle\Delta{\sf Q}^2_i(r=0,m=1,k)\rangle = 2\sum_n \langle c^2_{k,n}\rangle
						     R_{i,1,n}^2(0),\label{Qfluct}
\end{equation}
with the radial functions normalized, 
$\langle c^2_{k,n}\rangle=k_B T/L\xi(k^2+\lambda_r^{(n)})$ 
given by
equipartion,
and the factor 2 coming from the twofold degeneracy ($\cos\phi$ and $\sin\phi$ in the angular
part). The displacement $\langle{\bf u}_k^2\rangle$ of the line is obtained from (\ref{Qfluct})
by means of (\ref{displac}).
If only the lowest radial mode is taken into account in (\ref{Qfluct}), which is equivalent to
using the non-renormalized string model, 
the thermal amplitude is obviously underestimated, Fig.~\ref{amplituda}, or conversely,
the extracted value of $K$ is too low.
The error increases rather slowly (logarithmically) with $R$; for experimentally relevant
scales it is of the order of 100\%.
The failure of the string model is best demonstrated in the limit $R\to\infty$: 
for a fixed and nonzero $k$,
$\langle{\bf u}_k^2\rangle\to 0$ if only the lowest radial mode is taken into account, since
the norm (squared) of the mode is logarithmically diverging when $R\to\infty$. 
Alternatively, the same can be seen by noting the logarithmic divergence of the line tension
(\ref{F0}). 
\renewcommand{\subfigcapskip}{-0.2cm}
\renewcommand{\subfigtopskip}{0cm}
\renewcommand{\subfigbottomskip}{0cm}

Thus, to simulate or analyze a real observation of the disclination line fluctuations, e.g., 
by polarizing microscopy, for a given $k$ one has to sum up a proper number of radial modes 
possessing a considerable thermal amplitude; 
the maximum number is set by the resolution of the instrument ---
the summation should be stopped at the latest when the first
zero of the radial functions becomes comparable to the resolution. 
As in a real situation $1/k$ is large compared to the resolution, 
the series is always truncated earlier by the diminishing thermal amplitude and hence 
there is no cutoff ambiguity. The two characteristic regimes are illustrated in Fig.~{\ref{amplituda}}: a),
if $\pi/k\ll R$, many radial modes have to be summed up and the fluctuation amplitude is
essentially independent of the system size $R$, whereas b), for $\pi/k\ge R$ 
(severe confinement) a few modes 
suffice and the fluctuation amplitude is suppressed by decreasing $R$.

Of course, there exist other soft fluctuation modes, i.e., 
those with $m\ne 1$, 
which have not been focused on. Their radial eigenfunctions have zeros at $r=0$, and hence,
these modes are not important for the disclination line fluctuations, neither are they
particularly characteristic for the defect structure.
The $m=0$ soft mode, which involves only the component $x_{-1}$
(Eq.~(\ref{R-1}) is decoupled in this case), is a special case, 
corresponding to the Goldstone rotation of $\sf Q$ around the $z$ axis. Naturally, it
is present regardless of the configuration; the defect structure brings
about merely a decay of the eigenfunction at the uniaxial core.

Our results are valid in the approximation
where the free energy is invariant to separate rotations of space and the order parameter
(nematics and smectics-C in the one constant approximation, superfluids).
More precisely, we require that the order parameter gradients in the $z$ direction enter 
the free energy in the form of square terms only, Eq.(\ref{elast}).
If this symmetry is broken, the gradients along $z$ are in general 
coupled to those in the $xy$ plane. 
In a related manner, superconductors and their analogues,
smectics-A (SmA) \cite{degennes},
besides the $XY$-model (nematics, superfluids) degrees of freedom 
$\Psi = \vert\Psi\vert\exp{i\phi}$ 
(wave function of the superconducting state or the amplitude and phase of the smectic
modulation)
possess an additional vector order parameter
(vector potential $\bf A$ or the c-director, respectively). 
Due to the coupling between $\nabla\Psi$ and $\bf A$ described by the free energy density
term $\vert\left(-i\hbar\nabla-e{\bf A}\right)\Psi\vert^2$/2m, gradients in the $z$ direction
induce a change in $\bf A$ (analogously for SmA).
As a consequence, in the string fluctuations the cross-sectional
structure of the line defect (vortex or dislocation) is not merely displaced 
but also deformed, and, hence, the actual line tension is not simply the free energy per 
unit length (not even asymptotically). 
In this case one can define an effective line tension and must still
renormalize the string model with the contribution of higher radial modes.

This work was supported by the Slovenian Office of Science
(Program P1-0099) and US-Slovene NSF Joint Found 
(Grant No. 9815313).

\end{document}